\documentclass[conference, dvipsnames]{IEEEtran}
\IEEEoverridecommandlockouts
\usepackage[utf8]{inputenc}
\usepackage{amsmath, amssymb}
\usepackage{graphicx}
\usepackage[printonlyused,nolist]{acronym}
\usepackage{xcolor}
\usepackage{hyperref}
\usepackage{algorithm}
\usepackage{algpseudocode}
\usepackage{etoolbox}
\usepackage{multirow}
\usepackage{url}
\usepackage{tikz, pgfplots}
\usepackage{ifthen}
\usepackage{xfp}
\usepackage{gensymb}
\usepackage{LMSnodeshapes}
\usepackage{multirow}
\usepackage{LMScolor}
\usepackage{booktabs}
\usetikzlibrary{backgrounds,arrows}
\pgfplotsset{compat=1.15}
\usetikzlibrary{calc}
\usetikzlibrary{decorations}
\usetikzlibrary{snakes}
\usetikzlibrary{shapes}

\newboolean{drawEcho} 
\setboolean{drawEcho}{true} 
\newboolean{drawBgNoise} 
\setboolean{drawBgNoise}{true} 
\newboolean{drawNearEnd} 
\setboolean{drawNearEnd}{true} 
\newboolean{colorRoom} 
\setboolean{colorRoom}{false} 

\newboolean{drawKF} 
\setboolean{drawKF}{true} 
\newboolean{drawNL} 
\setboolean{drawNL}{false} 
\newboolean{drawAdaptation} 
\setboolean{drawAdaptation}{true} 
\newboolean{drawAir} 
\setboolean{drawAir}{false} 
\newboolean{drawPF} 
\setboolean{drawPF}{true} 

\colorlet{airEst}{black}
\colorlet{echoSig}{black}				
\colorlet{nearEndSig}{black}
\colorlet{bgNoiseSig}{black}
\colorlet{pfCol}{black}
\colorlet{nl}{black}		
\colorlet{adaptation}{black}	
\colorlet{aecBox}{black}

\pgfdeclarelayer{background}
\pgfdeclarelayer{foreground}
\pgfsetlayers{background,main,foreground}

\pgfplotsset{
	/pgfplots/layers/Bowpark/.define layer set={
		axis background,background,axis grid,main,axis ticks,axis lines,axis tick labels,
		axis descriptions,axis foreground
	}{/pgfplots/layers/standard},
}

\DeclareMathOperator*{\bw}{\boldsymbol{w}}

\DeclareMathOperator*{\bd}{\boldsymbol{d}}
\DeclareMathOperator*{\bx}{\boldsymbol{x}}

\DeclareMathOperator*{\be}{\boldsymbol{e}}

\DeclareMathOperator*{\by}{\boldsymbol{y}}

\DeclareMathOperator*{\bs}{\boldsymbol{s}}
\DeclareMathOperator*{\ba}{\boldsymbol{a}}
\DeclareMathOperator*{\herm}{\text{H}}
\DeclareMathOperator*{\trans}{\text{T}}

\DeclareMathOperator*{\bPsi}{\boldsymbol{\Psi}}

\DeclareMathOperator*{\bz}{\boldsymbol{z}}

\begin{acronym}
	\acro{ENR}{echo-to-noise ratio}
	\acro{SER}{signal-to-echo ratio}
	\acro{ERLE}{echo return loss enhancement}
	\acro{SNR}{signal-to-noise ratio}
	\acro{PESQ}{perceptual evaluation of speech quality}
	\acro{STOI}{short-time objective intelligibility}
	\acro{KF}{Kalman Filter}
	\acro{PBKF}{Partitioned-Block KF}
	\acro{PBAF}{Partitioned-Block Adaptive Filter}
	\acro{PBC}{Partitioned-Block Convolution}
	\acro{ML}{Maximum-Likelihood}
	\acro{EM}{expectation-maximization}
	\acro{AEC}{acoustic echo canceller}
	\acro{PF}{postfilter}
	\acro{GRU}{gated recurrent unit}
	\acro{DPF}{deep learning-based PF}
	\acro{MMSE}{Minimum Mean Square Error}
	\acro{STFT}{short-time Fourier transform}
	\acro{DSB}{delay-and-sum BF}
	\acro{PSD}{PSD}
	\acro{CPSD}{cross power spectral density}
	\acro{RIR}{room impulse response}
	\acro{FIR}{finite impulse response}
	\acro{PDF}{Probability Density Function}
	\acro{VSS}{Variable Step Size}
	\acro{DFT}{discrete fourier transform}
	\acro{DNN}{deep neural network}
	\acro{KL}{Kullback-Leibler}
	\acro{NER}{near-end-to-echo ratio}
	\acro{AF}{adaptive filter}
	\acro{EPC}{Echo Path Change}
	\acro{RTF}{relative transfer function}
	\acro{MVDR}{minimum variance distortionless response}
	\acro{BF}{beamformer}
	\acro{SIMO}{single-input-multiple-output}
	\acro{FR}{frequency response}
	\acro{FDAF}{frequeny-domain adaptive filter}
\end{acronym}

\def\BibTeX{{\rm B\kern-.05em{\sc i\kern-.025em b}\kern-.08em
    T\kern-.1667em\lower.7ex\hbox{E}\kern-.125emX}}

\newtoggle{long}
\toggletrue{long}

\begin{document}

\title{\vspace*{-.27cm}\commentTHd{Deep Learning-Based Joint Control of Acoustic Echo Cancellation, Beamforming and Postfiltering}}

\author{\IEEEauthorblockN{Thomas Haubner and Walter Kellermann}
	\IEEEauthorblockA{\textit{Multimedia Communications and Signal Processing,}
		\textit{Friedrich-Alexander-University Erlangen-N\"urnberg,}\\
		Cauerstr. 7, D-91058 Erlangen, Germany,
		e-mail: \texttt{\{\commentTHd{thomas.haubner,walter.kellermann}\}@fau.de}}}

\newcommand{\commentTHa}[1]{\textcolor{black}{#1}}
\newcommand{\commentTHb}[1]{\textcolor{black}{#1}}
\newcommand{\commentTHc}[1]{\textcolor{black}{#1}}
\newcommand{\commentTHd}[1]{\textcolor{black}{#1}}
\newcommand{\commentTHe}[1]{\textcolor{black}{#1}}

\newcommand{\commentTHpar}[1]{\textcolor{black}{#1}}

\maketitle

\begin{abstract}
	\commentTHc{We introduce a novel method for controlling the functionality of a hands-free speech communication device which comprises a model-based \ac{AEC}, \ac{MVDR} \ac{BF} and spectral \ac{PF}. While the \ac{AEC} removes the early echo component, the \ac{MVDR} \ac{BF} and \ac{PF} suppress the residual echo and background noise. As key innovation, we suggest to use a single \ac{DNN} to jointly control the adaptation of the various algorithmic components. This allows for rapid convergence and high steady-state performance in the presence of high-level interfering double-talk. End-to-end training of the \ac{DNN} using a time-domain speech extraction loss function avoids the design of individual control strategies.}
\end{abstract}


\section{Introduction}
\label{sec:intro} 
%
%
%
\let\thefootnote\relax\footnotetext{Accepted for \textit{European Signal Processing Conference (EUSIPCO)} 2022.}
Driven by the increased usage of hands-free \commentTHc{voice communication devices}, acoustic echo control has recently become again a highly active research field \cite{wang_weighted_2021, valin_low-complexity_2021, aec_challenge}. Most modern acoustic echo control algorithms can be classified into three classes: Model-based system identification algorithms (traditionally referred to as \acp{AEC}) \cite{enzner_frequency-domain_2006, malik_double_talk_2020}, pure deep learning-based spectral \acp{PF} \cite{dual_stage_aec} and a combination of both \cite{wang_weighted_2021, valin_low-complexity_2021, combAdFiltAndComValDPF}. While traditional \acp{AEC} generalize well to unknown acoustic environments and introduce no distortion to the desired near-end speech signal, they are inherently limited by their model assumptions, e.g., the filter length and non-linearity model \cite{haensler2004acoustic}. In contrast, pure deep learning-based \ac{PF} approaches avoid this limitation by treating acoustic echo control as an interference suppression problem. Yet, this comes at the cost of requiring large models to minimize the distorting effect of spectrally masking the near-end speaker \cite{dual_stage_aec}. 
Note that echo suppression is in this sense vastly more challenging in comparison to typical acoustic noise suppression tasks, due to the much smaller \ac{SER} in comparison to \ac{SNR}. This problem can be mitigated by enhancing the \ac{SER} by a model-based \ac{AEC} before \commentTHc{applying} a \ac{DNN}-based \ac{PF} for residual echo suppression. This fusion of traditional \acp{AEC} with deep learning-based \acp{PF} has recently been shown to be superior to their individual counterparts for a variety of challenging scenarios \cite{aec_challenge}. Yet, this benefit is only obtained if the \ac{AEC} is equipped with a sophisticated adaptation control which ensures a rapid convergence and robust steady-state performance \commentTHc{in the presence of double-talk}. For this, machine learning-based approaches \commentTHc{were shown} to be superior to their traditional counterparts \cite{dic_ad_control_haubner,kfNN_haubner, dnn_aec_haubner, ivry_deep_nodate}.

Besides spectral \ac{PF}-based approaches, \acp{BF} have shown great potential for achieving high echo and noise attenuation \cite{herbordt_joint_2005, luis_valero_2019, park_state-space_2019, cohen_online_2021}. They exploit the spatial diversity of the interfering signal components, i.e., echo and noise, and the desired near-end speaker by \commentTHc{linear spatiotemporal processing of} the microphone or the echo-reduced \ac{AEC} error signals. 
%
Yet, they require precise parameter \commentTHd{estimates of, e.g. the relative transfer functions and/or \ac{CPSD} of the interference,} to minimize the distortion of the near-end speech signal. This problem has been addressed by jointly estimating the \ac{AEC} and \ac{BF} weights by optimizing a least squares criterion \cite{herbordt_joint_2005}, online \acl{EM} algorithms \cite{park_state-space_2019,cohen_online_2021} and recently also \ac{DNN}-supported approaches \cite{carbajal_joint_2020, zhang_multi-channel_2021}.  However, whereas  \cite{zhang_multi-channel_2021} completely omits the model-based \ac{AEC} and \commentTHc{suggests} a \ac{DNN}-supported \ac{MVDR} \ac{BF} \cite{chakrabarty_timefrequency_2019} for echo and noise reduction, \cite{carbajal_joint_2020} considers a \commentTHe{convolutive} narrowband \ac{AEC} and is limited to offline applications. Furthermore, there have also been investigations of multi-microphone \ac{DNN}-only echo and noise {control} algorithms \cite{kothapally_joint_2021} which however require again large networks and are trained for specific \commentTHd{microphone} array topologies.

\commentTHc{In this paper we introduce a novel method for jointly controlling a model-based broadband \ac{AEC}, \ac{MVDR} \ac{BF} and spectral \ac{PF}} for online acoustic echo and noise \commentTHc{reduction}. While the \ac{AEC} improves the \ac{SER} by cancelling the early echo, the \ac{MVDR} \ac{BF} and \ac{PF} \commentTHc{exploit the spatial and \commentTHd{spectrotemporal} variability of the signal components to suppress the residual echo and noise.}
To achieve fast convergence and robust steady-state performance in the presence of \commentTHc{double-talk}, we suggest to use a single \ac{DNN} to jointly control the parameter \commentTHc{adaptation} of the algorithmic components. The \ac{DNN} \commentTHc{is trained} end-to-end with respect to a component-based \commentTHc{time-domain loss} function which allows to trade interference suppression against speech distortion. \commentTHc{This avoids the development of individual control strategies and allows the algorithmic components to synergistically interact.}

We use bold uppercase letters for matrices and bold lowercase letters for vectors with time-domain quantities being indicated by an underline. The $M\times M$-dimensional identity matrix\commentTHc{, all-zero matrix and \ac{DFT} matrix are denoted by $\boldsymbol{I}_M$, $\boldsymbol{0}_M$ and $\boldsymbol{F}_M$, respectively. Furthermore, we introduce the} element-wise product operator~$\odot$, the linear convolution operator~$*$ and the Euclidean norm $||\cdot||$. The transposition and Hermitian transposition are denoted by $(\cdot)^{\trans}$ and $(\cdot)^{\herm}$, respectively. Finally, we indicate the $m$th element of a vector by $\left[ \cdot \right]_m$.

\section{Signal Model}
\label{sec:sig_mod}
\begin{figure}[t!]
	\def\nearEndImageWidth{.038}
	\input{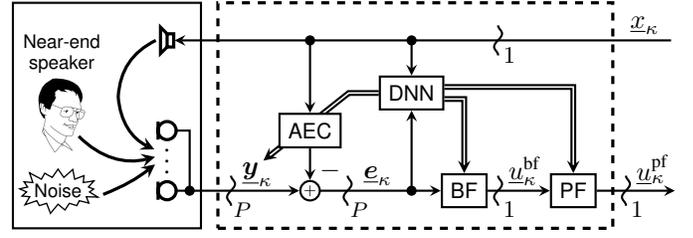}
	\vspace*{-.4cm}
	\caption{Block diagram of the proposed \commentTHc{\ac{DNN}-controlled} speech enhancement algorithm with the loudspeaker signal $\underline{x}_\kappa $, the microphone signal $\underline{\boldsymbol{y}}_\kappa $, the \commentTHc{multichannel} \ac{AEC} error signal \commentTHc{$\underline{\boldsymbol{e}}_\kappa $}, \commentTHc{the \ac{BF} output $\underline{{u}}_{ \kappa}^{\text{bf}} $} and the \ac{PF} output \commentTHc{$\underline{{u}}_{ \kappa}^{\text{pf}} $}. \commentTHe{The numbers below the curved lines correspond to the respective channel dimensions.}}
	\label{fig:alg_overview}
\end{figure}

We consider a \commentTHc{hands-free voice communication} scenario with $P$ microphones as shown in Fig.~\ref{fig:alg_overview}. The multichannel time-domain microphone signal \commentTHb{${\underline{\boldsymbol{y}}}_{\kappa}$} at sample index $\kappa$ is modelled as a linear superposition of an echo component ${\underline{\boldsymbol{d}}}_{\kappa}$, a speech component ${{\underline{\boldsymbol{s}}}}_{\kappa}$ and \commentTHc{a noise} component ${\underline{\boldsymbol{n}}}_{\kappa}$ as follows:
\begin{equation}
{\underline{\boldsymbol{y}}}_{\kappa} = {\underline{\boldsymbol{d}}}_{\kappa} + {\boldsymbol{\underline{s}}}_{\kappa} +  {\underline{\boldsymbol{n}}}_{\kappa} \in \mathbb{R}^P.
\label{eq:mic_sig_td}
\end{equation}
Furthermore, the echo and the speech images at the microphones are modelled by a linear convolution of the loudspeaker signal $\underline{x}_{\kappa}$ and the dry, i.e., non-reverberant, speech signal $\underline{s}_{\kappa}^{\text{dr}}$ with according \acp{RIR} $\underline{h}_{p,\kappa}$ and $\underline{g}_{p,\kappa} $
\begin{align}
\left[{\underline{\boldsymbol{d}}}_{\kappa}  \right]_p \commentTHc{= {\underline{{d}}}_{p,\kappa} }  &= \underline{h}_{p,\kappa} * \underline{x}_{\kappa}  \label{eq:echo_sig_td} \\
\left[ {\underline{\boldsymbol{s}}}_{\kappa} \right]_p \commentTHc{= {\underline{{s}}}_{p,\kappa} }  &= \underline{g}_{p,\kappa} * \underline{s}_{\kappa}^{\text{dr}} \label{eq:speech_sig_td}.
\end{align}
As we consider in the following a block-based online processing of the signals, we introduce the \commentTHc{time-domain loudspeaker signal block of length $M=2R$}
\begin{equation}
\commentTHc{\underline{\bx}_{\tau}^{\text{bl}} = \begin{pmatrix} {\underline{\bx}}_{\tau-1}^{\text{\commentTHc{in}}} \\ {\underline{\bx}}_{\tau}^{\text{\commentTHc{in}}}\end{pmatrix}\in \mathbb{R}^{M} }
\label{eq:td_block_x}
\end{equation}
%
which is composed of two innovation blocks \commentTHc{of the form}
\begin{equation}
\commentTHc{{\underline{\bx}}_{\tau}^{\text{\commentTHc{in}}} = \begin{pmatrix}\underline{x}_{\commentTHe{\tau} R - R + 1}, \underline{x}_{\commentTHe{\tau} R - R + 2}, \dots, \underline{x}_{\commentTHe{\tau} R}\end{pmatrix}^{{\text{T}}} \in \mathbb{R}^{R} }
\label{eq:innovation_block_x}
\end{equation}
with $R$ being the frameshift and $\tau$ the block index. 
\commentTHc{Analogously, we introduce the microphone innovation block \makebox{$\underline{\by}_{p,\tau}^{\text{in}} \in \mathbb{R}^R$} at the $p$th microphone.}	

\section{Proposed Online Acoustic Echo and Noise Control Algorithm}
\label{sec:ac_ec_and_noise_control}
%
%
\commentTHc{We first introduce the individual speech enhancement components, i.e., linear \ac{AEC}, \ac{MVDR} \ac{BF} and spectral \ac{PF}, before we turn to the \commentTHd{deep learning-based method for joint control.}}

\subsection{Acoustic Echo Cancellation}
\label{sec:aec}
%
%
\commentTHc{As \ac{AEC} we \commentTHc{consider} a straightforward extension of the popular \ac{FDAF} algorithm \cite{haykin_2002, shynk} to multiple microphones.} Here, an echo estimate at the $p$th microphone and block index $\tau$ is obtained by a linear convolution of the \ac{FIR} filter \makebox{$\hat{\underline{\boldsymbol{h}}}_{p,\tau}$} of length $R$ with the loudspeaker signal block $\commentTHc{\underline{\boldsymbol{x}}_{\tau}^{\text{bl}}}$ \commentTHc{(cf~Eq.~\eqref{eq:td_block_x})}. The linear convolution can efficiently be implemented by overlap-save processing in the \ac{DFT} domain
\begin{equation}
\widehat{{\underline{\bd}}}_{p,\tau}^{\text{\commentTHc{in}}} = \boldsymbol{Q}_1^{{\text{T}}} \boldsymbol{F}_M^{-1} \left(\left( \boldsymbol{F}_M \commentTHc{\underline{\boldsymbol{x}}_{\tau}^{\text{bl}}} \right) \odot \hat{\boldsymbol{h}}_{p,\tau} \right) \in \mathbb{R}^R
\label{eq:echo_estimate}
\end{equation}
with the \ac{DFT}-domain \ac{FIR} filter \makebox{$\hat{\boldsymbol{h}}_{p,\tau} = \boldsymbol{F}_M \boldsymbol{Q}_2 \hat{\underline{\boldsymbol{h}}}_{p,\tau} \in \mathbb{C}^M$} and the zero-padding matrix \makebox{$\boldsymbol{Q}_2^{\text{T}}= \begin{pmatrix}\boldsymbol{I}_{\commentTHb{R}} & \boldsymbol{0}_{\commentTHb{R}}\end{pmatrix}$}. Note that the constraint matrix \makebox{$\boldsymbol{Q}_1^{\text{T}} = \begin{pmatrix}\boldsymbol{0}_{\commentTHb{R}} & \boldsymbol{I}_R\end{pmatrix}$} ensures a linear convolution of the \ac{DFT}-domain product in brackets by discarding the elements corresponding to a circular convolution \cite{haykin_2002}. Subsequently, the time-domain error innovation block is computed by subtracting the echo estimate $\widehat{{\underline{\bd}}}_{p,\tau}^{\text{\commentTHc{in}}}$ \commentTHc{(cf.~Eq.~\eqref{eq:echo_estimate})} from the respective microphone observations $\underline{\by}_{p,\tau}^{\text{\commentTHc{in}}}$
\begin{equation}
\underline{{\be}}_{p,\tau}^{\text{\commentTHc{in}}} = \underline{{\by}}_{p,\tau}^{\text{\commentTHc{in}}} -\widehat{{\underline{{\bd}}}}_{p,\tau}^{\text{\commentTHc{in}}}.
\label{eq:error_comp_aec}
\end{equation}
%
%
\commentTHc{The filter coefficients $\hat{\boldsymbol{h}}_{p,\tau}$ are adapted} by the gradient-based update rule \cite{haykin_2002, shynk}
\begin{align}
\Delta \hat{\boldsymbol{h}}_{p,\tau}  &= \left( \boldsymbol{F}_M \commentTHc{\underline{\boldsymbol{x}}_{\tau}^{\text{bl}}} \right)^{*} \odot \commentTHc{\left(\boldsymbol{F}_M \boldsymbol{Q}_1 \underline{\boldsymbol{e}}_{p,\tau}^{{\text{\commentTHc{in}}}}\right)} \label{eq:grad_aec}\\
\hat{\boldsymbol{h}}_{p,\tau}  &= \hat{\boldsymbol{h}}_{p,\tau - 1} + \boldsymbol{Q}_3 \left(  \boldsymbol{\mu}_{p,\tau} \odot \Delta \hat{\boldsymbol{h}}_{p,\tau}\right) \label{eq:update_aec}
\end{align}
%
%
\commentTHc{with the frequency- and microphone-dependent step-size vector  $\boldsymbol{\mu}_{p,\tau}$.}
Furthermore, the gradient-constraint projection matrix \makebox{$\boldsymbol{Q}_3 = \boldsymbol{F}_M \boldsymbol{Q}_2 \boldsymbol{Q}_2^{\text{T}} \boldsymbol{F}_M^{-1}$} ensures that the estimate $\hat{\boldsymbol{h}}_{p,\tau}$ corresponds to a zero-padded \ac{FIR} filter $\hat{\underline{\boldsymbol{h}}}_{p,\tau}  = \boldsymbol{Q}_2^{\text{T}} \boldsymbol{F}_M^{-1} \hat{{\boldsymbol{h}}}_{p,\tau}$ \cite{shynk}. The performance of this update decisively depends on a proper choice of the step-size vector $\boldsymbol{\mu}_{p,\tau}$. Due to its rapid convergence speed and \commentTHc{double-talk} robustness, we use a straightforward extension of the \ac{DNN}-\ac{FDAF} approach \cite{dnn_aec_haubner} to multiple microphones. In particular, we suggest the time-varying step-size vector 
\begin{equation}
\commentTHc{\left[\boldsymbol{\mu}_{p,\tau} \right]_{f} = \frac{ {m}_{ f,\tau}^{\mu}}{ {\commentTHe{\Psi}}_{f,\tau}^{\text{XX}} + \frac{M}{R} \left| m_{f,\tau}^{\text{aec}} \commentTHc{\left[ \boldsymbol{F}_M \boldsymbol{Q}_1 \underline{\boldsymbol{e}}_{p,\tau}^{{\text{\commentTHc{in}}}}  \right]_f } \right|^2  }}
\label{eq:dnn_step_size}
\end{equation}
%
%
with ${\Psi}{}_{f,\tau}^{\text{XX}}$ being the loudspeaker signal \acs{PSD} and \makebox{${m}_{f,\tau}^{\mu} \in [0,1]$} and $m_{f,\tau}^{\text{aec}}  \in [0,1]$ \commentTHc{being auxiliary variables which are provided for each frequency bin $f$ and time $\tau$ by a \ac{DNN}} \commentTHc{(cf.~Sec.~\ref{sec:dnn_control})}. The step-size vector \eqref{eq:dnn_step_size} allows the \ac{DNN} to set entries entirely to zero, while in addition being robust to varying signal powers due to the frequency-selective loudspeaker \commentTHc{signal power} and error power normalization. In addition, the \ac{DNN} can eliminate the error power normalization to address the different reasons for large errors, i.e., interfering signal activity (\commentTHc{double-talk}) or system misalignment. The loudspeaker signal \acs{PSD} ${\Psi}{}_{f,\tau}^{\text{XX}} $ is estimated by recursive averaging
\begin{align}
{\Psi}{}_{f,\tau}^{\text{XX}} &= \lambda_{\text{X}} ~{\Psi}{}_{f,\tau-1}^{\text{XX}} + (1-\lambda_{\text{X}}) ~ \left| \left[\boldsymbol{F}_M \commentTHc{\underline{\boldsymbol{x}}_{\tau}^{\text{bl}}}\right]_f  \right|^2
\label{eq:inputSigPSDESt} 
\end{align}
with \commentTHc{$\lambda_{\text{X}}\in(0,1)$} being an easy-to-choose hyperparameter.

\subsection{MVDR Beamforming}
\label{sec:bf}
As the linear spatial filtering is conducted in the \ac{STFT} domain, we introduce the broadband \ac{STFT}-domain error signal at the $p$th microphone \commentTHc{(cf.~Eq.~\eqref{eq:error_comp_aec})}
\begin{equation}
\boldsymbol{e}_{p,\tau} = \boldsymbol{F}_M \commentTHc{\left( \boldsymbol{b} \odot \begin{pmatrix} {\underline{\be}}_{ \commentTHe{p,} \tau-1}^{\text{\commentTHc{in}}} \\ {\underline{\be}}_{\commentTHe{p,}\tau}^{\text{\commentTHc{in}}}\end{pmatrix} \right)} \in \mathbb{C}^M
\label{eq:stft_err_sig_def}
\end{equation}
%
%
%
%
%
with \commentTHc{$\boldsymbol{b}$ being a Hamming window.}
\commentTHd{By} concatenating the narrowband error signal components \commentTHb{(cf.~Eq.~\eqref{eq:stft_err_sig_def})} at the different microphones, we obtain the multichannel error signal vector
\begin{equation}
\tilde{\boldsymbol{e}}_{f,\tau} = \begin{pmatrix}
\left[{\boldsymbol{e}}_{1,\tau}  \right]_f & \dots & \left[{\boldsymbol{e}}_{P,\tau}  \right]_f
\end{pmatrix}^{\trans}  \in \mathbb{C}^P
\label{eq:mult_error_signal}
\end{equation}
at frequency \commentTHc{bin} $f$.
Note that in the following all \commentTHc{\ac{STFT}-domain narrowband signals} are indicated by a tilde. 

The \commentTHc{single-channel} \ac{STFT}-domain output of the \ac{MVDR} \ac{BF}
\begin{equation}
\commentTHc{\tilde{u}_{ f,\tau}^{\text{bf}}} = \tilde{\bw}_{f,\tau}^{\herm} \tilde{\boldsymbol{e}}_{f,\tau}
\label{eq:bf_out}
\end{equation}
is computed as the inner product of the multichannel error signal $\tilde{\boldsymbol{e}}_{f,\tau}$ (cf.~Eq.~\eqref{eq:mult_error_signal}) and the \ac{MVDR} weight vector \cite{van_trees}
\begin{equation}
\tilde{\bw}_{f ,\tau} = \frac{ \left( {\tilde{\bPsi}_{f,\tau}^{\text{ZZ}}} + \delta_1 \boldsymbol{I}_{\commentTHb{P}} \right)^{-1} \tilde{\ba}_{f,\tau} }{ \tilde{\ba}_{f,\tau}^{\text{H}} \left( {\tilde{\bPsi}_{f,\tau}^{\text{ZZ}}} + \delta_1 \boldsymbol{I}_{\commentTHb{P}} \right)^{-1} \tilde{\ba}_{f,\tau} + \delta_2 } \label{eq:mvdr_result}
\end{equation}
%
%
%
%
\commentTHc{with $\delta_{1}$ and $\delta_2$ being regularization constants. The interfering signal \ac{CPSD} matrix \makebox{$\tilde{\bPsi}{}_{f,\tau}^{\text{ZZ}} \in \mathbb{C}^{P \times P} $} and near-end \ac{RTF} vector \makebox{$\tilde{\ba}_{f,\tau} \in \mathbb{C}^{ P}$} are computed from \commentTHd{estimates of the} interference, i.e., residual echo and noise, and speech image}
\begin{align}
\commentTHc{\left[ \commentTHc{\hat{\tilde{{\bz}}}_{f,\tau}}  \right]_p }&= \commentTHc{\left(1-m_{p,f,\tau}^{\text{bf}} \right) \left[ \tilde{\boldsymbol{e}}_{f, \tau} \right]_p },
\label{eq:noise_sig_est} \\
\commentTHc{{\left[\hat{\tilde{{\bs}}}_{f,\tau} \right]_p}}  &= \commentTHc{ m_{p,f,\tau}^{\text{bf}} \left[ \tilde{\boldsymbol{e}}_{f ,\tau} \right]_p},
\label{eq:speech_sig_est} 
\end{align}
\commentTHd{respectively,} \commentTHc{with $m_{p,f,\tau}^{\text{bf}} \in [0,1]$ being a microphone-, frequency- and time-dependent mask that is provided by a \ac{DNN} (cf.~Sec.~\ref{sec:dnn_control}). While the interfering signal \ac{CPSD} matrix $\tilde{\bPsi}{}_{f,\tau}^{\text{ZZ}}$ is directly calculated by recursively averaging the outer products of the interference estimates $ \commentTHc{\hat{\tilde{{\bz}}}_{f,\tau}}$ (cf.~Eq.~\eqref{eq:noise_sig_est}) }
\begin{equation}
\commentTHc{\tilde{\bPsi}_{f,\tau}^{\text{ZZ}} =\lambda_{\text{Z}} ~\tilde{\bPsi}_{f,\tau-1}^{\text{ZZ}} + (1-\lambda_{\text{Z}}) ~  \hat{\tilde{{\bz}}}_{f,\tau} \hat{\tilde{{\bz}}}_{f,\tau}^{\herm}},
\label{eq:noise_psd} 
\end{equation}
\commentTHc{the near-end \ac{RTF} vector $\tilde{\ba}_{f,\tau}$ is computed as normalized eigenvector corresponding to the maximum eigenvalue of the speech \ac{CPSD} matrix estimate}
\begin{equation}
\tilde{\bPsi}_{f,\tau}^{\text{SS}} =\lambda_{\text{S}} ~\tilde{\bPsi}_{f,\tau-1}^{\text{SS}} + (1-\lambda_{\text{S}}) ~ {\hat{\tilde{{\bs}}}_{f,\tau}  \hat{\tilde{{\bs}}}_{f,\tau}^{\herm}}.
\label{eq:speech_psd} 
\end{equation}
A computationally \commentTHc{efficient solution to the} eigenvector problem is given by the power iteration algorithm \cite{GoluVanl96}
\begin{align}
\tilde{\ba}_{f,\tau} &= \tilde{\bPsi}_{f,\tau}^{\text{SS}} \tilde{\ba}_{f,\tau-1}  \label{eq:pow_it} \\
\tilde{\ba}_{f,\tau} &=  \frac{ \tilde{\ba}_{f,\tau} }{  \left[ \tilde{\ba}_{f,\tau} \right]_1 } \label{eq:pow_it_norm}
\end{align}
\commentTHc{\commentTHd{whose simplicity allows} for a numerically robust end-to-end training of the \ac{DNN} parameters (cf.~Sec.~\ref{sec:dnn_control}) \cite{boeddeker_convolutive_2021}.}
Note that\commentTHc{, in Eq.~\eqref{eq:pow_it_norm}, we use the first microphone as reference \commentTHd{without loss of generality}.}

\begin{figure}[htbp]
	 \vspace*{-.2cm}
	\centering
		
	\begin{tikzpicture}[node distance=2.0cm, >=stealth]

	\node (featSig) at (0,0) {};
	\node [rectangle, draw, thick, right of=featSig] (ffIn) { {$f_{\text{FF}}^{\text{in}}$} };
	\node [rectangle, draw, thick, right of=ffIn] (ffGru) {$f_{\text{GRU}}$};
	\node [rectangle, draw, thick, above right of=ffGru, yshift=-0.1cm, xshift=.5cm] (ffOutMu) {{$f_{\text{FF}}^{\mu}$}};
	\node [rectangle, draw, thick, above right of=ffGru, yshift=-1.0cm, xshift=.5cm] (ffOutE) {{$f_{\text{FF}}^{\text{aec}}$}};
	\node [rectangle, draw, thick, below right of=ffGru, yshift=1.0cm, xshift=.5cm] (ffOutBf) {{$f_{\text{FF}}^{\text{bf}}$}};
		\node [rectangle, draw, thick, below right of=ffGru, yshift=.1cm, xshift=.5cm] (ffOutPf) {{$f_{\text{FF}}^{\text{pf}}$}};
	\node [right of=ffOutMu] (maskMu) {};
	\node [right of=ffOutE] (maskE) {};
	\node [right of=ffOutBf] (maskBf) {};
		\node [right of=ffOutPf] (maskPf) {};
	
	\draw [thick, ->] (featSig) -- (ffIn.west) node [midway, above] {${\boldsymbol{v}}_{\tau}$};
	\draw [thick, ->] (ffIn.east) -- (ffGru.west);
	\draw [thick, ->] (ffGru.east) to [out=0,in=180] (ffOutMu.west);
	\draw [thick, ->] (ffGru.east) to [out=0,in=180] (ffOutE.west);
	\draw [thick, ->] (ffGru.east) to [out=0,in=180] (ffOutBf.west);
	\draw [thick, ->] (ffGru.east) to [out=0,in=180] (ffOutPf.west);
	\draw [thick, ->] (ffOutMu.east) -- (maskMu) node [midway, above]  {${m}_{f, \tau}^{{\mu}}$}; 
	\draw [thick, ->] (ffOutE.east) -- (maskE) node [midway, above]  {${m}_{f,\tau}^{{\text{aec}}}$};
	\draw [thick, ->] (ffOutBf.east) -- (maskBf) node [midway, above] {${m}_{p, f,\tau}^{\text{bf}}$}; 
\draw [thick, ->] (ffOutPf.east) -- (maskPf) node [midway, above] {${m}_{ f,\tau}^{\text{pf}}$};
	
	\draw [thick, ->] ($(ffGru.north)+(.3,0)$) to [out=75,in=0] ($(ffGru.north)+(.0,.4)$) to [out=180,in=105] ($(ffGru.north)+(-.3,0)$) ;
	\node (delay) at ($(ffGru.north)+(.0,.6)$)  {$\tau-1$};

	
	\end{tikzpicture}
	\caption{\commentTHc{\ac{DNN} architecture for jointly controlling the \ac{AEC}, \ac{BF} and \ac{PF}}.}
	\label{fig:nn_architecture}
	 \vspace*{-.15cm}
\end{figure}
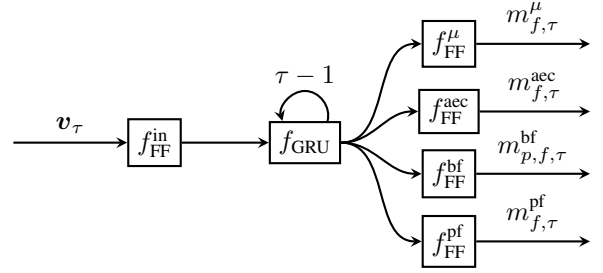

\subsection{Spectral Postfilter}
\label{sec:scpf}
The \ac{MVDR} \ac{BF} output $\commentTHc{\tilde{u}_{ f,\tau}^{\text{bf}}}$ (cf.~Eq.~\eqref{eq:bf_out}) is subsequently processed by a scalar spectral \ac{PF} as follows
\begin{equation}
\commentTHc{\tilde{u}_{ f,\tau}^{\text{pf}} = m_{f,\tau}^{\text{pf}} \tilde{u}_{ f,\tau}^{\text{bf}}}
\label{eq:scpf_out}
\end{equation}
with \commentTHc{$m_{f,\tau}^{\text{pf}}$ being a frequency-dependent mask which is inferred by a \ac{DNN} (cf.~Sec.~\ref{sec:dnn_control}).}

\subsection{Deep Learning-Based Adaptation Control}
\label{sec:dnn_control}
In the following, we will describe how to \commentTHc{determine the masks ${m}_{f,\tau}^{{\mu}}$ and ${m}_{f,\tau}^{{\text{aec}}}$ (cf.~\eqref{eq:dnn_step_size}), ${m}_{p,f,\tau}^{\text{bf}}$ (cf.~\eqref{eq:noise_sig_est} and \eqref{eq:speech_sig_est}) and ${m}_{f,\tau}^{\text{pf}}$ (cf.~\eqref{eq:scpf_out}) which control the adaptation of the \ac{AEC}, \ac{MVDR} \ac{BF} and \ac{PF}, respectively, by a single \ac{DNN}.}
As \commentTHc{input feature vector \commentTHc{$\boldsymbol{v}_\tau$} for the \ac{DNN}} we use a concatenation of the logarithmic magnitude spectrum of the \commentTHc{\ac{STFT}-domain loudspeaker signal and} the \ac{AEC} error signals at the different microphones \makebox{$\boldsymbol{e}_{1,\tau}$, $\dots$, $\boldsymbol{e}_{P,\tau}$} (cf.~Eq.~\eqref{eq:stft_err_sig_def}). The feature elements are \commentTHc{normalized} by estimating their mean and standard deviation during training. Note that due to the symmetry of the \ac{STFT}-domain signals we only use \commentTHc{the frequency} components up to $F=\frac{M}{2}+1$ with $M$ being the even \ac{DFT} length. 

The \ac{DNN} is composed of a feed-forward layer, which condenses the input feature vector \commentTHc{$\boldsymbol{v}_\tau$} to a lower dimension \commentTHc{$Q$}, two stacked \ac{GRU} layers, which extract temporal information, and finally four different feed-forward layers with sigmoid activation which map to the various \commentTHc{masks as shown in Fig.~\ref{fig:nn_architecture}}. The architecture is \commentTHc{chosen to yield after the \ac{GRU} layer a condensed representation of the convergence state of the \ac{AEC}, the noise spectrum and the near-end \commentTHd{source} activity.}
The \ac{DNN} parameters $\boldsymbol{\theta}$ are trained end-to-end w.r.t. the component-based loss function
\begin{equation}
	{\mathcal{J}(\boldsymbol{\theta})} =   \alpha~ || \text{pf}({\underline{\boldsymbol{d}}}_{1:K}) ||  + \beta~|| \text{pf}({\underline{\boldsymbol{n}}}_{1:K})||  +   || \underline{s}_{1:K}^{\text{ref}} - \text{pf}({\underline{\boldsymbol{s}}}_{1:K}) ||
	\label{eq:cost_fct}
\end{equation}
with $\text{pf}({\underline{\boldsymbol{d}}}_{1:K})$, $\text{pf}({\underline{\boldsymbol{n}}}_{1:K})$ and  $\text{pf}({\underline{\boldsymbol{s}}}_{1:K})$ denoting \commentTHc{the length-$K$ time-domain signal of the echo, noise and speech images, respectively, after being processed by the algorithm.}
While the first two terms in \eqref{eq:cost_fct} quantify the \commentTHc{suppression of} interference, i.e., echo and noise, the last term represents the distortion of the near-end speech signal w.r.t. a time-domain reference signal $\underline{s}_{\kappa}^{\text{ref}}$.
The trade-off between interference suppression and near-end distortion can be \commentTHc{controlled} by the hyperparameters $\alpha$ and $\beta$.
\commentTHc{Eq. \eqref{eq:cost_fct}} is an echo-aware time-domain version of the frequency\commentTHb{-}domain noise suppression loss proposed in \cite{xia_weighted_2020}.
\commentTHc{Note that the end-to-end optimization of the \ac{DNN}} avoids the design of desired oracle target masks and instead directly evaluates their effect on the speech extraction performance.

\section{Experiments}
\label{sec:experiments}
We will now evaluate the proposed algorithm for a variety of different multi-microphone acoustic echo and noise control scenarios. Each scenario comprises a circular microphone array with \commentTHpar{four} uniformly-distributed elements and a random diameter in the range \commentTHpar{$[7 \text{cm},~15 \text{cm} ]$}. The array is placed randomly in a shoebox room with random \commentTHc{lengths, widths and heights} in the ranges \commentTHpar{$ [3 \text{m},~8 \text{m}]$, $ [3 \text{m},~8 \text{m}]$ and $[2.0 \text{m},~3.5 \text{m}]$}\commentTHc{, respectively,} and random reverberation time \commentTHpar{$T_{60}\in[0.2 \text{s},~0.6 \text{s}]$}. The loudspeaker and near-end speaker positions are sampled from the \commentTHpar{azimuthal angle range $[0 \degree,~360 \degree]$, elevation angle range $[-20 \degree,~20 \degree]$ and distance ranges  $[0.1 \text{m},~0.5 \text{m}]$ and  $[0.5\text{m},~2.0\text{m}]$, respectively.}   
All \acp{RIR} are simulated by the image method \cite{allen1979image, habets2010room} with a filter length of \commentTHpar{$\commentTHc{\text{max}}\{6000, \lfloor f_s T_{60} \rfloor \}$} and sampling frequency $f_s=16$ kHz. The \commentTHd{dry} near-end speaker and loudspeaker signals $\underline{s}_\kappa^{\text{dr}}$ and $\underline{x}_\kappa$ (cf.~Eq.~\eqref{eq:mic_sig_td}), respectively, were drawn from different subsets of the \textit{LibriSpeech} dataset \cite{libri_speech} comprising \commentTHpar{$143$} speakers each. The near-end speaker activity started randomly in the range \commentTHpar{$[1 \text{s},~4 \text{s}]$} which allows to individually evaluate \commentTHc{single-talk} and \commentTHc{double-talk} performance of the algorithm. We consider recorded single-channel background noise \commentTHe{signals} from cafés, street junctions, public transportation and pedestrian areas \cite{chime} which were made spherically diffuse by using \cite{spat_coh_noise}.
The near-end speech and background noise signals were scaled to create a random echo-to-near-end and echo-to-noise ratio in the ranges \commentTHpar{$[-10 \text{dB},~10 \text{dB}]$ and $[10 \text{dB},~25 \text{dB}]$}, respectively.

The frame shift $R$ and block length $M$ were chosen as \commentTHpar{$1024$ and $2048$}, respectively, which results in an \ac{AEC} filter length of \commentTHpar{$1024$} taps. 
%
Furthermore, the recursive \acs{PSD} averaging factors were set to \commentTHpar{$\lambda_{\text{X}}=0.5$, $\commentTHc{\lambda_{\text{Z}}=\lambda_{\text{S}}}=0.99$ and the \ac{MVDR} regularization constants were chosen as $\delta_{1}=\delta_{2}=0.01$.} The condensed feature dimension \commentTHc{$Q$} of the \ac{DNN} was set to \commentTHpar{$256$} which results in overall \commentTHpar{$3.9$ million} parameters. As reference signal \commentTHc{$s_{\kappa}^{\text{ref}}$} in the cost function \eqref{eq:cost_fct}, we use \commentTHpar{a delay-and-sum \ac{BF} applied to the near-end speech image $\underline{\boldsymbol{s}}_{\kappa}$, and} \commentTHc{the weighting parameters were chosen as $\alpha=\beta=1$.} The network was trained by using the {ADAM} optimizer with a step-size of \commentTHpar{$0.001$} and \commentTHpar{$1.4$h} of training data. For evaluating the algorithms we simulated an additional \commentTHpar{$8.3$min ($50$ scenarios with $10$s each)} which were disjoint from the training data, i.e., different signals, arrays and environments.

\begin{figure}[tbp]
	\centering
	\newlength\fwidth
	\setlength\fwidth{.88\columnwidth}
	\iftoggle{long}{%
		\input{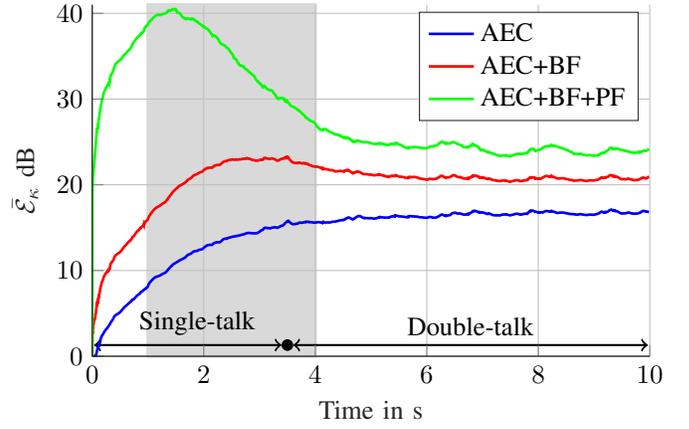}	
	}{%
		\vspace*{4cm}
		
	}
	
	\vspace*{-.5 cm}
	\caption{Average time-dependent \ac{ERLE} $\bar{\mathcal{E}}_{\kappa}$ of the various algorithmic components. The starting point of the \commentTHc{near-end} speaker activity is indicated by a black dot and sampled randomly from the shaded range.}
	\label{fig:onlineEchoSuppPerf}
	\vspace*{-.249 cm} 
\end{figure}

To evaluate the convergence rate and steady-state echo suppression performance of the various \commentTHc{algorithmic} components, we introduce the time-dependent logarithmic \ac{ERLE} obtained at the first microphone
\begin{equation}
{\mathcal{E}_{\kappa}} = 10  \log_{10}\frac{ \hat{\mathbb{E}} \left[   | \underline{d}_{1,\kappa} |^2\right] }{ \hat{\mathbb{E}} \left[  | \commentTHe{ \text{pr}({\underline{\boldsymbol{d}}}_{\kappa})} |^2 \right] } 
\end{equation}
with $\hat{\mathbb{E}}$ denoting recursive averaging. 
Furthermore, $\text{pr}({\underline{\boldsymbol{d}}}_{\kappa})$ denotes the processed echo image ${\underline{\boldsymbol{d}}}_{\kappa}$ at the output of the various speech enhancement components, i.e., \ac{AEC}, \ac{MVDR} \ac{BF} and \ac{PF}.
\commentTHe{Note that $\text{pr}(\cdot)$ provides a scalar output because we select the first channel of the \ac{AEC} and the outputs of the \ac{BF} and the \ac{PF} are anyway scalar.}
Fig.~\ref{fig:onlineEchoSuppPerf} shows the average time-dependent \ac{ERLE} $\bar{\mathcal{E}}_{\kappa}$ over $50$ different scenarios. The starting point of the \commentTHc{near-end} speaker activity is indicated by a black dot and sampled randomly from the shaded range. We conclude from Fig.~\ref{fig:onlineEchoSuppPerf} that the adaptive \ac{AEC} and \ac{MVDR} \ac{BF} converge despite the interfering \commentTHc{double-talk} rapidly to its steady-state. The \ac{PF} drastically increases the echo suppression during the \commentTHc{single-talk} period at the beginning. Yet, after the \commentTHc{near-end} speaker starts talking, the \ac{PF} attenuation is significantly reduced to minimize the distortion of the desired speech signal. 

We will now investigate the echo and noise suppression performance of the algorithmic components during \commentTHc{single-talk} and \commentTHc{double-talk} and relate it to the induced distortion of the \commentTHc{near-end} speech signal. For this we introduce the logarithmic \ac{ERLE} and \commentTHc{the} noise suppression factor at the first microphone
\begin{align}
		\mathcal{E} &= 10  \log_{10} \frac{ \sum_{\kappa=\commentTHe{N_1}}^{\commentTHe{N_2}} | \underline{d}_{1,\kappa} |^2}{ \sum_{\kappa=\commentTHe{N_1}}^{\commentTHe{N_2}} | \commentTHe{ \text{pr}({\underline{\boldsymbol{d}}}_{\kappa})} |^2} \\
	\mathcal{N} &= 10  \log_{10} \frac{ \sum_{\kappa=\commentTHe{N_1}}^{\commentTHe{N_2}} | \underline{n}_{1,\kappa} |^2}{ \sum_{\kappa=\commentTHe{N_1}}^{\commentTHe{N_2}} | \text{pr}({\underline{\boldsymbol{n}}}_{\kappa}) |^2}\commentTHc{,}
\end{align}
\commentTHc{respectively,} with $\text{pr}(\cdot)$ denoting again the processed signal in brackets at the output of the algorithmic processing chain. \commentTHe{Note that choosing the summation bounds $N_1$ and $N_2$ allows to assess different time periods, e.g., single- or double-talk.}
To evaluate the near-end speech distortion \commentTHc{during double-talk} we use the PESQ (Perceptual Evaluation of Speech Quality \cite{pesq})
\begin{equation}
	{\mathcal{S}}_{\text{dist}} = \text{pesq}(\commentTHc{\underline{s}_{\commentTHe{N_1:N_2}}^{\text{ref}}}, \text{pr}(\commentTHc{\underline{\boldsymbol{s}}_{\commentTHe{N_1:N_2}}}))
\end{equation}
of the clean reference signal \commentTHc{$\underline{s}_{\commentTHe{N_1:N_2}}^{\text{ref}}$} and the processed speech image \commentTHc{$ \text{pr}(\underline{\boldsymbol{s}}_{\commentTHe{N_1:N_2}})$}. In addition, to assess the general performance during \commentTHd{double-talk} we introduce the PESQ
\begin{equation}
{\mathcal{S}} = \text{pesq}(\commentTHc{\underline{s}_{\commentTHe{N_1:N_2}}^{\text{ref}}}, \text{pr}(\commentTHc{\underline{\boldsymbol{x}}_{\commentTHe{N_1:N_2}}}))
\end{equation}
with $\text{pr}(\commentTHc{\underline{\boldsymbol{x}}_{\commentTHe{N_1:N_2}}})$ being the near-end speech estimate after the various speech enhancement components. All performance measures have been averaged over \commentTHpar{$50$} different scenarios (\commentTHpar{$8.3$min}) which is denoted by an overbar. 
Tab.~\ref{tab:perfEvalTab} shows the average performance measures during \commentTHc{single-talk} and \commentTHc{double-talk} in comparison to an oracle baseline, i.e., using the first $R$ coefficients of the true \acp{RIR} $\underline{h}_{p,\kappa} $ as \ac{AEC}, the oracle \commentTHc{signals} for \commentTHc{computing the \ac{CPSD} matrices of} the \ac{MVDR} \ac{BF} and the oracle magnitude \commentTHd{ratio} mask as spectral \ac{PF}. We conclude that the cancellation of the early echo by the \commentTHc{\ac{DNN}-controlled} \ac{AEC} results in a PESQ gain of $0.63$ \commentTHb{which is close to the oracle baseline}. The subsequent \ac{MVDR} \ac{BF} increases the PESQ by additional $0.29$. This improvement is obtained by the increased echo and noise suppression while introducing only minimal distortion to the \commentTHc{near-end} speech signal (cf.~$\bar{\mathcal{S}}_{\text{dist}}$ for \commentTHc{double-talk} and \ac{MVDR} \ac{BF}). The spectral \ac{PF} achieves almost no additional PESQ gain during double-talk as the improved interference suppression is traded against increased speech distortion. Yet, the \ac{PF} obtains remarkable echo and noise suppression during the \commentTHc{single-talk} period at the beginning. This is of particular importance due to the limited performance of the \ac{AEC} and \ac{MVDR} {\ac{BF}} during the initial convergence phase. Note that the partly overlapping convergence phase and \commentTHc{single-talk} period explain also the different interference suppression performance of the \ac{AEC} and \ac{MVDR} \ac{BF} during \commentTHc{single-talk} and \commentTHc{double-talk}. \commentTHc{Finally, as the average runtime of the proposed algorithm for processing one signal innovation block of duration \makebox{$64$ms} on an \textit{IntelXeon CPU-E3-1275 v6@3.80GHz} is \makebox{$3$ms}, we confirm real-time capability on such platforms.}
\begin{table}[tbp]
	\vspace*{-.05cm}
	\caption{Average performance measures of the various speech enhancement components during \commentTHc{single-talk} and \commentTHc{double-talk} in comparison to an oracle baseline.}
	\vspace*{-.15cm}
		\setlength{\tabcolsep}{5.8pt}
	\begin{center}
		\begin{tabular}{l l || c c | c c c  c}
			\toprule
			\multicolumn{2}{c||}{Algorithm }			& \multicolumn{2}{c}{\commentTHd{Single-talk}}  & \multicolumn{4}{c}{\commentTHd{Double-talk}} \\ 
			&		& $\bar{\mathcal{E}}$ &  $\bar{\mathcal{N}}$  & $\bar{\mathcal{E}}$ &  $\bar{\mathcal{N}}$ & $\bar{\mathcal{S}}_{\text{dist}}$ & $\bar{\mathcal{S}}$      \\ \midrule
			\multicolumn{2}{c||}{Unprocessed}	  	& \textemdash 	& \textemdash		& \textemdash 	& \textemdash		& \textemdash & $1.20$   \\ \midrule
			\multirow{3}{*}{\commentTHc{Proposed}} \hspace*{-.5cm} 	& AEC 	& $\hphantom{x}9.8$ 	& \textemdash		& $16.5$ 	& \textemdash		&\textemdash & $1.83$   \\ 
			& \commentTHc{AEC+BF} 						& $17.9$ 	& $\hphantom{x}8.9$		& $20.8$ 	& $5.7$		& $3.57$ & $2.12$   \\  
			& \commentTHc{AEC+BF+PF} 							& $37.9$ 	& $26.3$				& $23.8$ 	& $8.0$		& $3.00$ & $2.13$   \\
			\midrule 
			\multirow{3}{*}{Oracle} & AEC 	& \commentTHb{$19.9$} 	&\textemdash					& \commentTHb{$19.6$} 	& \textemdash		& \textemdash & \commentTHb{$1.90$}   \\ 
			& \commentTHc{AEC+BF} 						& \commentTHb{$29.1$} 	& \hspace*{.14cm}\commentTHb{$9.7$}					& \commentTHb{$28.5$} 	& \hspace*{.14cm}\commentTHb{$6.4$}		& \commentTHb{$3.48$} & \commentTHb{$2.35$}   \\  
			& \commentTHc{AEC+BF+PF} 							& \commentTHb{$60.3$} 	& \commentTHb{$36.9$}					& \commentTHb{$34.2$} 	& \commentTHb{$12.3$}		& \commentTHb{$3.46$} & \commentTHb{$3.22$}   \\ 
			\bottomrule 
		\end{tabular} 
	\end{center}
	\label{tab:perfEvalTab}
	\vspace{-.3cm}  
\end{table}

\section{Conclusion}
\label{sec:summaryOutlook}
In this paper, we introduced a novel online acoustic echo and noise \commentTHc{reduction} algorithm \commentTHd{with a \ac{DNN}-based joint control of acoustic echo cancellation, \ac{MVDR} beamforming and postfiltering as most distinctive feature.} \commentTHc{The proposed algorithm achieves rapid convergence and robust steady-state performance in the presence of high-level interfering double-talk.}
%
%
%
%
%

\bibliographystyle{IEEEbib}
{\small\bibliography{refs}}

\begin{thebibliography}{10}

\bibitem{wang_weighted_2021}
Z.~Wang et~al.,
\newblock ``Weighted recursive least square filter and neural network based
  residual echo suppression for the aec-challenge,''
\newblock in {\em Int. Conf. Acoust., Speech, Signal Process.}, Toronto, CA,
  June 2021, pp. 141--145.

\bibitem{valin_low-complexity_2021}
J.~Valin et~al.,
\newblock ``Low-{Complexity}, {Real}-{Time} {Joint} {Neural} {Echo} {Control}
  and {Speech} {Enhancement} {Based} {On} {Percepnet},''
\newblock in {\em Int. Conf. Acoust., Speech, Signal Process.}, Toronto, CA,
  June 2021, pp. 7133--7137.

\bibitem{aec_challenge}
K.~Sridhar et~al.,
\newblock ``{ICASSP} 2021 {A}coustic {E}cho {C}ancellation {C}hallenge:
  {D}atasets, {T}esting {F}ramework, and {R}esults,''
\newblock in {\em Int. Conf. Acoust., Speech, Signal Process.}, Toronto, CA,
  June 2021, pp. 151--155.

\bibitem{enzner_frequency-domain_2006}
G.~Enzner and P.~Vary,
\newblock ``Frequency-domain adaptive {Kalman} filter for acoustic echo control
  in hands-free telephones,''
\newblock {\em Signal Process.}, vol. 86, no. 6, pp. 1140--1156, June 2006.

\bibitem{malik_double_talk_2020}
S.~Malik et~al.,
\newblock ``Double-talk {Robust} {Multichannel} {Acoustic} {Echo}
  {Cancellation} using {Least}-{Squares} {MIMO} {Adaptive} {Filtering}:
  {Transversal}, {Array}, and {Lattice} {Forms},''
\newblock {\em IEEE Trans. Signal Process.}, 2020.

\bibitem{dual_stage_aec}
N.~L. Westhausen and B.~T. Meyer,
\newblock ``Acoustic echo cancellation with the dual-signal transformation lstm
  network,''
\newblock in {\em Int. Conf. Acoust., Speech, Signal Process.}, Toronto, CA,
  June 2021, pp. 7138--7142.

\bibitem{combAdFiltAndComValDPF}
Mhd.~M. Halimeh et~al.,
\newblock ``Combining adaptive filtering and complex-valued deep postfiltering
  for acoustic echo cancellation,''
\newblock in {\em Int. Conf. Acoust., Speech, Signal Process.}, Toronto, CA,
  June 2021, pp. 121--125.

\bibitem{haensler2004acoustic}
E.~H{\"a}nsler and G.~Schmidt,
\newblock {\em Acoustic {E}cho and {N}oise {C}ontrol: {A} practical
  {A}pproach},
\newblock Wiley-Interscience, NJ, USA, 2004.

\bibitem{dic_ad_control_haubner}
T.~Haubner et~al.,
\newblock ``Noise-robust adaptation control for supervised acoustic system
  identification exploiting a noise dictionary,''
\newblock in {\em Int. Conf. Acoust., Speech, Signal Process.}, Toronto, CA,
  June 2021, pp. 945--949.

\bibitem{kfNN_haubner}
T.~Haubner et~al.,
\newblock ``A {S}ynergistic {K}alman- and {D}eep {P}ostfiltering {A}pproach to
  {A}coustic {E}cho {C}ancellation,''
\newblock in {\em European Signal Process. Conf.}, Dublin, IR, Aug. 2021, pp.
  990--994.

\bibitem{dnn_aec_haubner}
T.~Haubner et~al.,
\newblock ``{E}nd-to-end deep learning-based adaptation control for
  frequency-domain adaptive system identification,''
\newblock in {\em Int. Conf. Acoust., Speech, Signal Process.}, Singapore, SG,
  May 2022, pp. 766--770.

\bibitem{ivry_deep_nodate}
A.~Ivry et~al.,
\newblock ``Deep adaptation control for acoustic echo cancellation,''
\newblock in {\em Int. Conf. Acoust., Speech, Signal Process.}, Singapore, SG,
  May 2022, pp. 741--745.

\bibitem{herbordt_joint_2005}
W.~Herbordt et~al.,
\newblock ``Joint {Optimization} of {LCMV} {Beamforming} and {Acoustic} {Echo}
  {Cancellation} for {Automatic} {Speech} {Recognition},''
\newblock in {\em Int. Conf. Acoust., Speech, Signal Process.}, Philadelphia,
  USA, March 2005.

\bibitem{luis_valero_2019}
M.~L. Valero and E.~A.~P. Habets,
\newblock ``Low-{Complexity} {Multi}-{Microphone} {Acoustic} {Echo} {Control}
  in the {Short}-{Time} {Fourier} {Transform} {Domain},''
\newblock {\em IEEE Audio, Speech, Language Process.}, vol. 27, no. 3, pp.
  595--609, Mar. 2019.

\bibitem{park_state-space_2019}
J.~Park and J.-H. Chang,
\newblock ``State-{Space} {Microphone} {Array} {Nonlinear} {Acoustic} {Echo}
  {Cancellation} {Using} {Multi}-{Microphone} {Near}-{End} {Speech}
  {Covariance},''
\newblock {\em IEEE Audio, Speech, Language Process.}, vol. 27, no. 10, pp.
  1520--1534, Oct. 2019.

\bibitem{cohen_online_2021}
N.~Cohen et~al.,
\newblock ``An online algorithm for echo cancellation, dereverberation and
  noise reduction based on a {Kalman}-{EM} {Method},''
\newblock {\em Eurasip J. Audio Speech Music Process.}, vol. 2021, no. 1, Dec.
  2021.

\bibitem{carbajal_joint_2020}
G.~Carbajal et~al.,
\newblock ``Joint {NN}-{Supported} {Multichannel} {Reduction} of {Acoustic}
  {Echo}, {Reverberation} and {Noise},''
\newblock {\em IEEE Audio, Speech, Language Process.}, vol. 28, pp. 2158--2173,
  2020.

\bibitem{zhang_multi-channel_2021}
H.~Zhang and D.~Wang,
\newblock ``Multi-{Channel} and {Multi}-{Microphone} {Acoustic} {Echo}
  {Cancellation} {Using} {A} {Deep} {Learning} {Based} {Approach},''
\newblock in {\em Annual Conf. of the Int. Speech Comm. Assoc.}, Brno, Czechia,
  Aug. 2021.

\bibitem{chakrabarty_timefrequency_2019}
S.~Chakrabarty and E.~A.~P. Habets,
\newblock ``Time–{Frequency} {Masking} {Based} {Online} {Multi}-{Channel}
  {Speech} {Enhancement} {With} {Convolutional} {Recurrent} {Neural}
  {Networks},''
\newblock {\em IEEE J. Sel. Top. Signal Process.}, vol. 13, no. 4, pp.
  787--799, Aug. 2019.

\bibitem{kothapally_joint_2021}
V.~Kothapally et~al.,
\newblock ``Joint {AEC} {and} {Beamforming} with {Double}-{Talk} {Detection}
  using {RNN}-{Transformer},''
\newblock {\em arXiv preprint:2111.04904}, 2021.

\bibitem{haykin_2002}
S.~Haykin,
\newblock {\em Adaptive {F}ilter {T}heory},
\newblock Prentice Hall, NJ, USA, 2002.

\bibitem{shynk}
J.~J. Shynk,
\newblock ``Frequency-domain and multirate adaptive filtering,''
\newblock {\em IEEE Signal Process. Mag.}, vol. 9, no. 1, pp. 14--37, 1992.

\bibitem{van_trees}
H.~L.~Van Trees,
\newblock {\em Optimum Array Processing: Part {IV} of Detection, Estimation,
  and Modulation Theory},
\newblock Wiley, 2002.

\bibitem{GoluVanl96}
G.~H. Golub and C.~F. Van~Loan,
\newblock {\em Matrix Computations},
\newblock The Johns Hopkins University Press, third edition, 1996.

\bibitem{boeddeker_convolutive_2021}
C.~Boeddeker et~al.,
\newblock ``Convolutive {Transfer} {Function} {Invariant} {SDR} {Training}
  {Criteria} for {Multi}-{Channel} {Reverberant} {Speech} {Separation},''
\newblock in {\em Int. Conf. Acoust., Speech, Signal Process.}, Toronto, CA,
  June 2021, pp. 8428--8432.

\bibitem{xia_weighted_2020}
Y.~Xia et~al.,
\newblock ``Weighted {Speech} {Distortion} {Losses} for
  {Neural}-{Network}-{Based} {Real}-{Time} {Speech} {Enhancement},''
\newblock in {\em Int. Conf. Acoust., Speech, Signal Process.}, Barcelona, ES,
  May 2020, pp. 871--875.

\bibitem{allen1979image}
J.-B. Allen and D.~A. Berkley,
\newblock ``Image method for efficiently simulating small-room acoustics,''
\newblock {\em J. Acoust. Soc. Am.}, vol. 65, no. 4, pp. 943--950, 1979.

\bibitem{habets2010room}
E.~A.~P. Habets,
\newblock ``Room {I}mpulse {R}esponse {G}enerator,''
\newblock Tech. {R}ep., Technische Universiteit Eindhoven, Sept. 2010.

\bibitem{libri_speech}
V.~{Panayotov} et~al.,
\newblock ``Librispeech: An {ASR} corpus based on public domain audio books,''
\newblock in {\em Int. Conf. Acoust., Speech, Signal Process.}, Brisbane, AU,
  Apr. 2015, pp. 5206--5210.

\bibitem{chime}
J.~Barker et~al.,
\newblock ``The third 'chime' speech separation and recognition challenge:
  Dataset, task and baselines,''
\newblock in {\em IEEE Autom. Speech Recognit. Underst. Workshop}, Scottsdale,
  USA, Dec 2016, pp. 504--511.

\bibitem{spat_coh_noise}
E.~A.~P. Habets et~al.,
\newblock ``{Generating} nonstationary multisensor signals under a spatial
  coherence constraint,''
\newblock {\em J. Acoust. Soc. Am.}, vol. 124, pp. 2911--2917, 2008.

\bibitem{pesq}
{ITU-T Recommendation P.862.2},
\newblock ``Wideband extension to recommendation {P}.862 for the assessment of
  wideband telephone networks and speech codecs,''
\newblock Recommendation, {ITU}, Nov. 2007.

\end{thebibliography}

\end{document}